\documentclass[journal = nalefd,manuscript=letter,layout=traditional]{achemso}

\setkeys{acs}{etalmode=truncate}
\setkeys{acs}{maxauthors=10}
\setkeys{acs}{articletitle=false}

\usepackage{amssymb}

\usepackage{graphicx}% Needed to include png/jpg/pdf figure files
\usepackage{amsmath}
\usepackage{bm}
\usepackage{color}
\usepackage{soul}

\title{Probing the inter-layer exciton physics in a MoS$_2$/MoSe$_2$/MoS$_2$ van der Waals heterostructure}

\author{M.\ Baranowski}
\affiliation{Laboratoire National des Champs Magn\'etiques Intenses, CNRS-UGA-UPS-INSA, 143 avenue de Rangueil, 31400 Toulouse, France}
\alsoaffiliation{Department of Experimental Physics, Faculty of Fundamental Problems of Technology, Wroclaw University of Science and Technology, Wybrzeze Wyspianskiego 27, 50-370 Wrocław, Poland}
\altaffiliation{These authors contributed equally to this work}
\author{A.\ Surrente}
\affiliation{Laboratoire National des Champs Magn\'etiques Intenses, CNRS-UGA-UPS-INSA, 143 avenue de Rangueil, 31400 Toulouse, France}
\altaffiliation{These authors contributed equally to this work}
\author{L.\ Klopotowski}
\affiliation{Institute of Physics, Polish Academy of Sciences, al.\ Lotnikow 32/46, 02-668 Warsaw, Poland}
\author{J.\ M.\ Urban}
\affiliation{Laboratoire National des Champs Magn\'etiques Intenses, CNRS-UGA-UPS-INSA, 143 avenue de Rangueil, 31400 Toulouse, France}
\author{N.\ Zhang}
\affiliation{Laboratoire National des Champs Magn\'etiques Intenses, CNRS-UGA-UPS-INSA, 143 avenue de Rangueil, 31400 Toulouse, France}
\author{D. K.\ Maude}
\affiliation{Laboratoire National des Champs Magn\'etiques Intenses, CNRS-UGA-UPS-INSA, 143 avenue de Rangueil, 31400 Toulouse, France}
\author{K.\ Wiwatowski}
\affiliation{Faculty of Physics, Astronomy and Informatics, Nicolaus Copernicus University, Grudziadzka 5, 87-100 Torun, Poland}
\author{S.\ Mackowski}
\affiliation{Faculty of Physics, Astronomy and Informatics, Nicolaus Copernicus University, Grudziadzka 5, 87-100 Torun, Poland}

\author{Y.\ C.\ Kung}
\affiliation{Electrical Engineering Institute and Institute of Materials Science and Engineering, \'Ecole Polytechnique F\'ed\'erale de Lausanne, CH-1015 Lausanne, Switzerland}

\author{D.\ Dumcenco}
\affiliation{Electrical Engineering Institute and Institute of Materials Science and Engineering, \'Ecole Polytechnique F\'ed\'erale de Lausanne, CH-1015 Lausanne, Switzerland}
\altaffiliation{present address Department of Quantum Matter Physics, Universit\'é de Gen\`{e}ve, 24 quai Ernest Ansermet, CH-1211, Geneva, Switzerland}

\author{A.\ Kis} \affiliation{Electrical Engineering Institute and Institute of Materials Science and Engineering, \'Ecole Polytechnique F\'ed\'erale de Lausanne, CH-1015 Lausanne, Switzerland}

\author{P.\ Plochocka}\email{paulina.plochocka@lncmi.cnrs.fr}
\affiliation{Laboratoire National des Champs Magn\'etiques Intenses, CNRS-UGA-UPS-INSA, 143 avenue de Rangueil, 31400 Toulouse, France}


\begin{document}

\begin{abstract}
Stacking atomic monolayers of semiconducting transition metal dichalcogenides (TMDs) has emerged as an effective way to engineer
their properties. In principle, the staggered band alignment of TMD heterostructures should result in the formation of 
inter-layer excitons with long lifetimes and robust valley polarization. However, these features have been observed simultaneously only in MoSe$_2$/WSe$_2$ heterostructures. Here we report on the observation of long lived inter-layer exciton emission in a
MoS$_2$/MoSe$_2$/MoS$_2$ trilayer van der Waals heterostructure. The inter-layer nature of the observed transition is confirmed
by photoluminescence spectroscopy, as well as by analyzing the temporal, excitation power and temperature dependence of
the inter-layer emission peak. The observed complex photoluminescence dynamics suggests the presence of quasi-degenerate
momentum-direct and momentum-indirect bandgaps. We show that circularly polarized optical pumping results in long lived valley
polarization of inter-layer exciton. Intriguingly, the inter-layer exciton photoluminescence has helicity opposite to the
excitation. Our results show that through a careful choice of the TMDs forming the van der Waals heterostructure it is possible
to control the circular polarization of the inter-layer exciton emission.
\end{abstract}

\begin{tocentry}

	\includegraphics[width=1\linewidth]{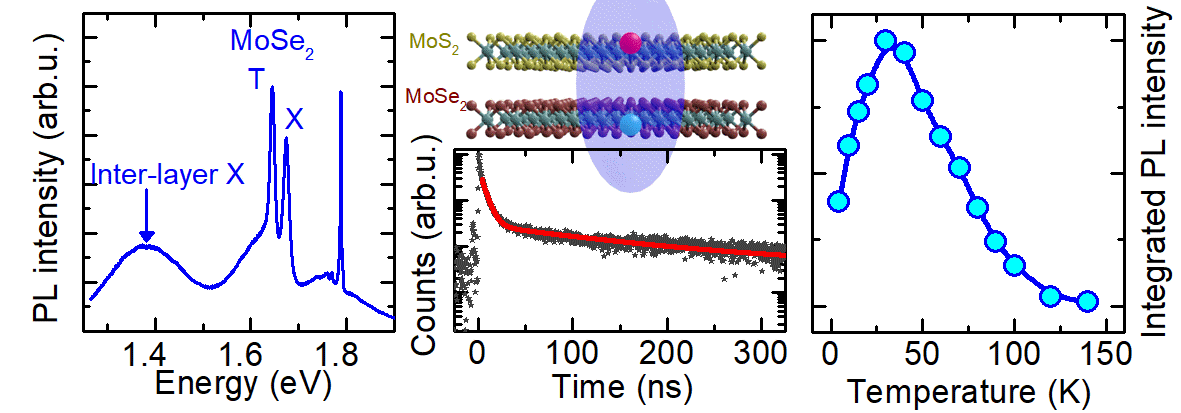}

\end{tocentry}

\maketitle

\section{Introduction}
Semiconductor heterostructures are at the heart of modern semiconductor science and technology. In addition to their wide spread
potential applications in every-day electronic and optoelectronic devices they also provide a rich playground for solid state
physics. The concept of engineering material properties via building heterostructures can be also applied to the recently
isolated monolayer van der Waals crystals. The vertical stacking of transition metal dichalcogenides (TMDs), boron nitride or
graphene \cite{Gong2014,Raja2017,Yang2013,Roy2013,Wang2015,Yu2012} permits the fabrication of a wide variety of heterostructures,
mainly because of the relaxed constraint regarding lattice matching. The weak van der Waals interaction between the layers
also allows to tune their electronic properties, not only by stacking different materials, but also by controlling their relative
orientation \cite{Yu2015, Nayak2017}.

Theoretical calculations predict a type II band alignment for van der Waals heterostructures based on semiconducting TMDs
\cite{Kang2013,zhang2016systematic}, which leads to an extremely efficient spatial separation of photocreated electron--hole pairs
\cite{Hong2014,Ceballos2014, Chen2016}. These carriers form inter-layer excitons, with significantly extended lifetimes compared
to their intra-layer counterparts \cite{Rivera2015,Rivera2016}. Consequently, TMD heterostructures are very promising for
photodetection and energy harvesting applications \cite{fur14,ros17}, or studies of bosonic properties of excitons, such as
Bose-Einstein condensation or superfluidity \cite{fog14}. In addition, inter-layer excitons can retain information about the
helicity of the exciting photons \cite{Rivera2016}, \textit{i.e.}, the valley selection rules of the individual layers are
reflected in the inter-layer emission. Since single layer TMDs suffer from very short exciton lifetimes and rapid valley
depolarization \cite{Lagarde2014,Wang2014, Robert2016,Wang_urbaszek2015,Zhu2014} increasing the recombination and depolarization
time is a crucial step towards the realization of the TMD based spin and valley devices. This makes van der Waals
heterostructures potentially useful for the practical realization of the valleytronic concepts and thus a profound understanding
of their physical properties is of paramount importance.

So far, type II band alignment and the expected charge transfer have been shown for a number of van der Waals heterostructures
such as WS$_2$/MoS$_2$ \cite{Gong2014,Hong2014,Rigosi2015,Heo2015,Chen2016}, WS$_2/$MoSe$_2$ \cite{Ceballos2015,Kozawa2016},
\cite{Fang2014,Chiu2015,Chiu2014}, WSe$_2$/WS$_2$\cite{Wang2016Hetero}, MoS$_2$/MoSe$_2$ \cite{Ceballos2014,
Kim2016,Surrente2017}. To date, inter-layer exciton emission has been observed mostly in ``mixed'' heterostructures containing
both tungsten and molybdenum based TMDs and only very recently emission of inter-layer exciton from MoSe$_2$/MoS$_2$
\cite{Mouri2017} heterostructure has been reported. However, a robust circular polarization has been observed only in
WSe$_2$/MoSe$_2$ heterobilayers\cite{Rivera2016}. Importantly, most of the investigations have been performed on mechanically
exfoliated TMD heterostructures. Despite their excellent optical properties, this approach based on the manual stacking of flakes
results in a very low yield of heterostructures (typically a few on a chip scale).

%However, the inter-layer exciton emission has been observed only for ``mixed'' heterostructures containing both tungsten and molybdenum based TMDs and long life times (in the range of ns) of the inter-layer exciton emission and robust circular polarization have been observed only in WSe$_2$/MoSe$_2$ heterobilayers\cite{Rivera2016}. To date, most of the investigations have been performed on mechanically exfoliated TMD heterostructures. Despite their excellent optical properties, this approach based on the manual stacking of flakes results in a very low yield of heterostructures (typically a few on a chip scale).

In this work, we demonstrate long lived inter-layer exciton emission from MoS$_2$/MoSe$_2$/MoS$_2$ trilayer van der Waals
heterostructures. Our fabrication approach relies on large area monolayer films deposited on a full chip scale via chemical vapor
deposition (CVD) \cite{dumcenco2015large,mitioglu2016magnetoexcitons,Surrente2017} and on the transfer of a multitude of single
layer flakes over areas of $\sim 100$\,$\mu$m$ \times 100$\,$\mu$m \cite{Surrente2017}. This paradigm enables us to fabricate a
large number of heterostructures, which can be characterized separately to post-select those displaying the best optical properties. PL mapping of our samples shows the area where the inter-layer excitons is observed can be as large as 20 $\mu$m $\times$ 20 $\mu$m and in a transfer zone of the size 100 $\mu$m $\times$ 100 $\mu$m, interlayer emission is particularly strong in several spots (see supplementary information fig.\ S1).  
Our approach represents an important step towards the fabrication of scalable van der Waals heterostructures for device integration. 
The inter-layer nature of the observed transition is confirmed by the enhanced photoluminescence (PL) under
resonant excitation of the MoSe$_2$ and MoS$_2$ A-excitons, as well as its temporal and excitation power. Comprehensive temperature and temporal measurements of the observed emission reveal complex recombination dynamics of the inter-layer, indicating on the presence of quasi-degenerate momentum-direct and momentum-indirect bandgaps.
Under circularly polarized excitation, the inter-layer exciton emission is intriguingly counter polarized; the emitted light has
the opposite helicity as compared to the excitation. Our results show that a careful choice of the TMDs forming the van der Waals
heterostructure allows to control the circular polarization of inter-layer exciton emission. This unexpected phenomenon gives an
additional degree of freedom for tailoring the properties of van der Waals heterostructures.

\section{Inter-layer exciton emission}

\begin{figure}[t]
\centering
\includegraphics[width=0.7\linewidth]{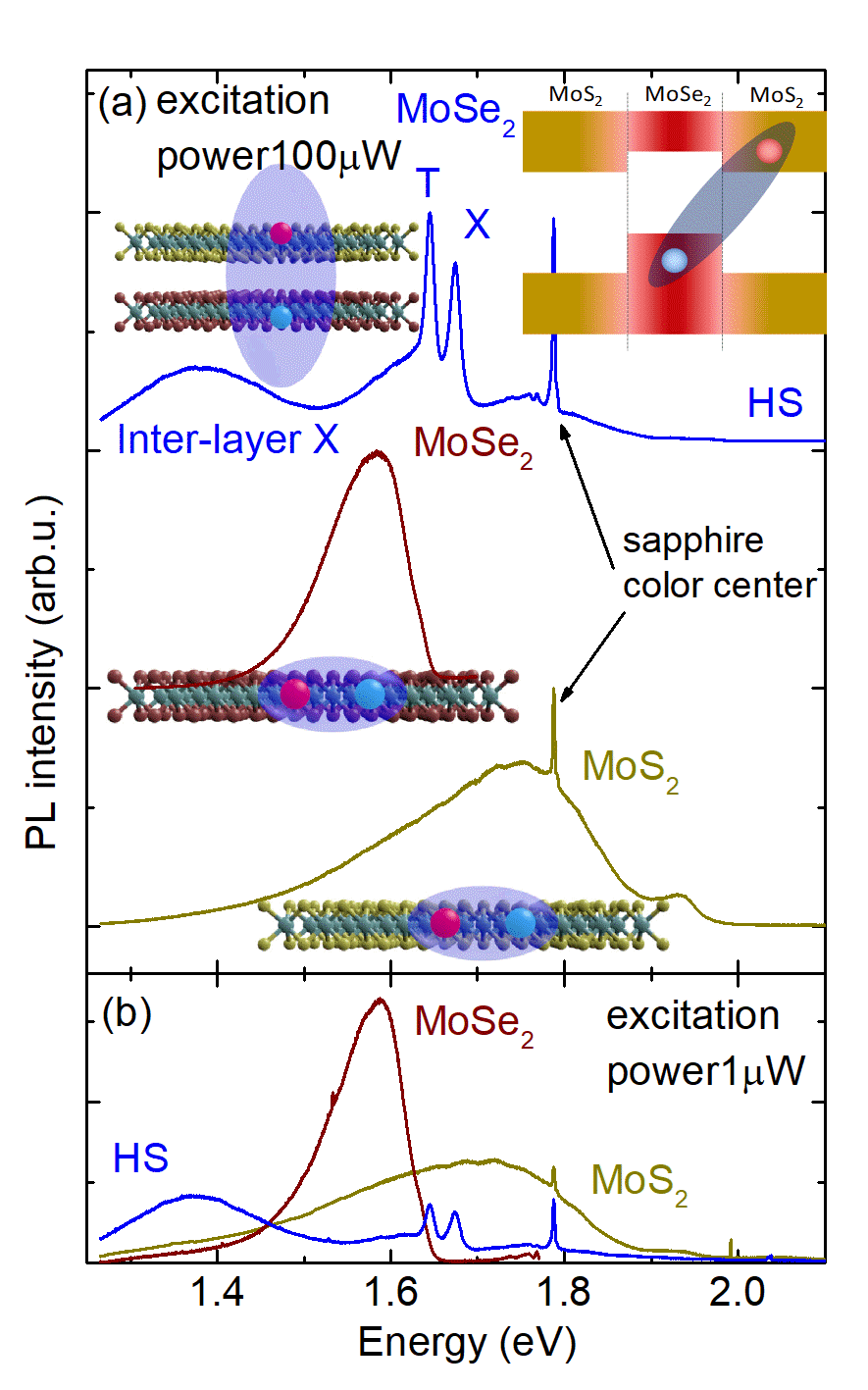}
\caption{ (a) Normalized photoluminescence spectra at $T=10$\,K of the trilayer structure and as-grown MoSe$_2$  and MoS$_2$
single layers. (b) Comparison of PL intensity from \emph{as grown} MoS$_2$ and MoSe$_2$ monolayers and heterostructure}
\label{fig:fig1}

\end{figure}

The trilayer MoS$_2$/MoSe$_2$/MoS$_2$ heterostructures used for this investigation consist of transferred single layer MoSe$_2$
flakes sandwiched between CVD grown single layers of MoS$_2$ \cite{Surrente2017} (see the micro-graph S2 presented in the
supplementary information). Representative PL spectra measured at $T=10$\,K for the trilayer structure are shown in
figure\,\ref{fig:fig1} (a) (spectra from other heterostructures are presented in fig. S3) together with the PL of
as-grown MoS$_2$ and MoSe$_2$. The PL of the as-grown layers is dominated by a broad, bound exciton emission, characteristic for
CVD-grown TMDs \cite{Tongay2013,Chang2014}. The optical properties of the trilayer stack are dramatically improved, with narrow,
well-resolved neutral and charged free exciton emission from MoSe$_2$ (see topmost spectrum in figure \ref{fig:fig1}(a)). The
bound exciton emission of MoSe$_2$ is strongly suppressed, consistent with the recently proposed defect healing scenario
\cite{Surrente2017}. The significant enhancement of optical quality of MoSe$_2$ layer achieved in our trilayer stack allows us to observe relatively weak emission at 1.38 eV, significantly below the trion and exciton lines of MoSe$_2$. %Here, we focus on the rather broad emission which appears around 1.38\,eV (below the MoS$_2$ and MoSe$_2$ exciton ground states). 
This low energy PL is observed only in the transfer zone (see also figure S4) and its presence is
accompanied by a significant suppression (by more than an order of magnitude) of the PL related to intra-layer excitons in
MoS$_2$ or MoSe$_2$ (see figure \ref{fig:fig1}(b)). Since this low energy PL is not observed in as grown MoS$_2$ or MoSe$_2$
mono-layers (fig. S3 (a)), a defect-related origin is unlikely. We therefore attribute this PL to the recombination of an
inter-layer exciton, with the hole localized in the MoSe$_2$ layer and the electron in the MoS$_2$ layer. Below, we provide
further experimental evidence supporting the identification of the low energy PL with the recombination of the inter-layer
exciton.

%The low energy emission is observed only in heterostructures, and is accompanied by a significant drop of MoSe$_2$ and MoS$_2$ PL intensity (about one and two orders of magnitude respectively), as shown in figure \ref{fig:fig1}(b). We assign this low energy peak to the radiative recombination of  inter-layer excitons (X$_\mathrm{I}$), formed by a hole localized in the MoSe$_2$ layer and an electron in one of the MoS$_2$ layers (see inset in figure \ref{fig:fig1}(a)). As the low energy emission is not observed in \emph{as grown} MoSe$_2$ or MoS$_2$, a defect related origin is unlikely.

\begin{figure*}[t]
\centering
\includegraphics[width=1\linewidth]{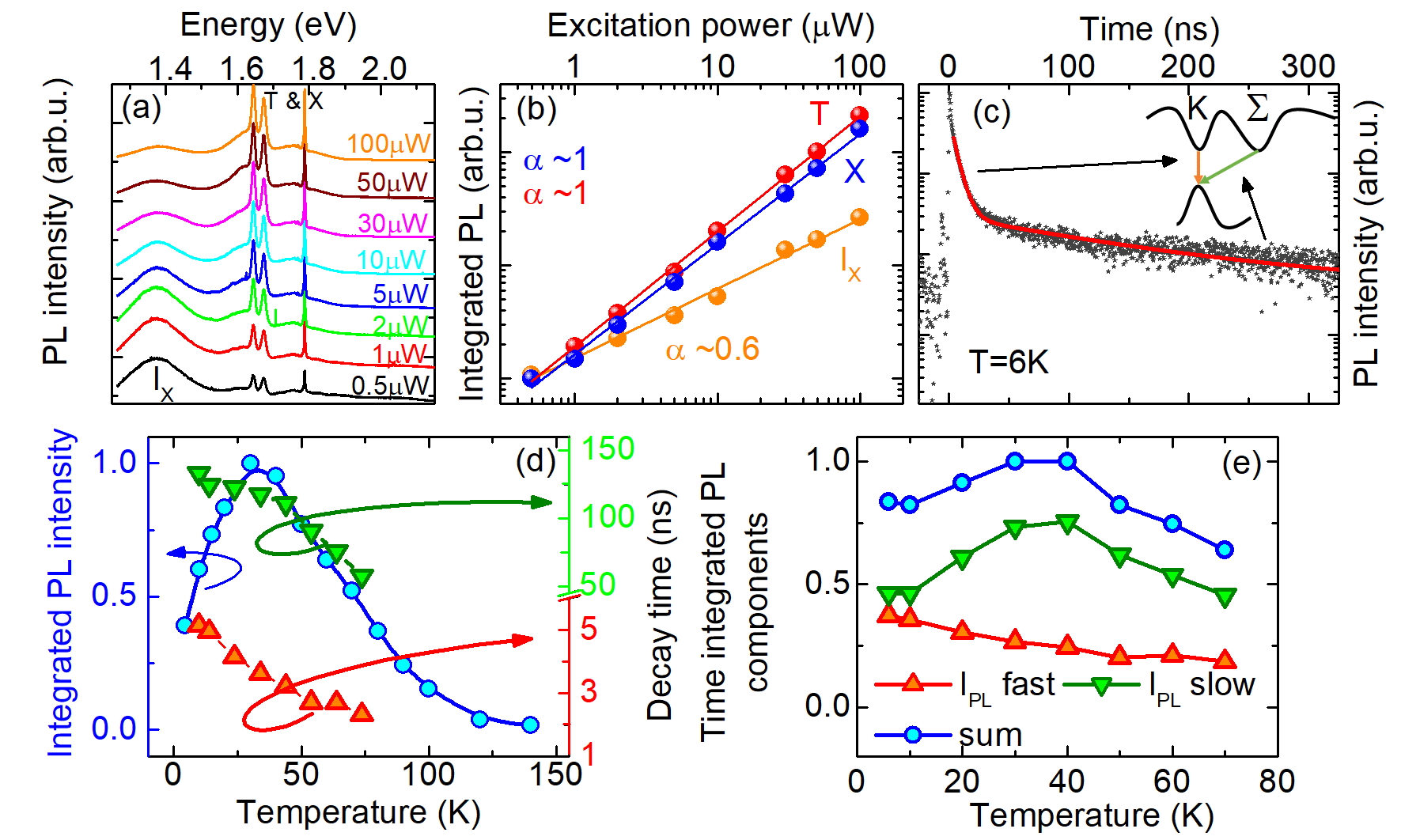}
\caption{(a) Power dependent PL of MoSe$_2$ from the trilayer stack (532\,nm cw laser) at $T=10$\,K. (b) Integrated PL intensity
of trion, exciton and inter-layer exciton PL as a function of excitation power. (c) Inter-layer exciton PL decay. The red
curve is the best fit to a bi-exponential decay model. A pulsed laser emitting at 1.937\,eV was used for the excitation. The inset schematically shows band alignment that can lead to observed bi-exponential PL kinetics.  (d)
Temperature dependence of the inter-layer PL intensity (blue points) and of the fast and slow decay components (orange and
green triangles, respectively). (e) Temperature dependence of time-integrated fast and slow PL decay components together
with their sum.} \label{fig:fig2}
\end{figure*}

Figure\,\ref{fig:fig2}(a) shows PL spectra measured at 10\,K at different excitation powers. The inter-layer exciton (I$_{\text{X}}$) displays a
significantly different power dependence as compared to the intra-layer exciton of MoSe$_2$. At low excitation power, the
I$_\mathrm{X}$ emission dominates the PL spectrum. With increasing power, the A-exciton intensity increases faster than that of
I$_\mathrm{X}$. In figure\,\ref{fig:fig2}(b) we plot the integrated intensity $I$ of the PL versus excitation power and we fit
this data to a power law ($I \propto P^\alpha$). The intensity of both the exciton and the trion increases linearly with the excitation power ($\alpha
\simeq 1$). The PL intensity of I$_\mathrm{X}$ increases sublinearly versus power ($\alpha \simeq 0.6$), which is a
characteristic feature of inter-layer transitions in van der Waals heterostructures \cite{Rivera2015, Rivera2016}. This results from the much longer inter-layer exciton lifetime ($\geq$ ns), compared to the spatially direct exciton transitions ($\simeq$
ps) and charge transfer times (few hundreds fs \cite{Ceballos2014, Chen2016}). Since only excitons situated in the light cone can recombine radiatively, the very effective charge transfer (as compared to the recombination process of inter-layer exciton) can saturate the optically active states, which leads to a sublinear dependence of PL intensity of the inter-layer exciton as a function of the excitation power. The long lifetime of I$_\mathrm{X}$ is confirmed by %inter-layer character of I$_\mathrm{X}$ emission is confirmed by 
the time resolved PL measurements presented in
figure\,\ref{fig:fig2}(c). The PL decay is characterized by two time scales, which can be extracted using a bi-exponential fit
(red line), giving decay times $\simeq 5$\,ns and $\simeq 135$\,ns for the fast and slow components of the decay at $T=6$\,K.

These two different time scales are consistent with the calculated band structure of MoSe$_2$/WSe$_2$ \cite{Nayak2017, miller2017long} and
MoS$_2$/WS$_2$ \cite{Heo2015} heterobilayers. These articles demonstrate that for coherent and incoherent orientation of the flakes, quasi-degenerate direct and indirect transitions in $k$-space appear in the band structure of the heterostructures, resulting in a bi-exponential decay, where the fast component of decay curve corresponds to the momentum-direct transition and the slow one is related to momentum-indirect transition (see schematics in figure \ref{fig:fig2}(c)). The simultaneous presence of momentum-direct and momentum-indirect transitions is also confirmed by the temperature dependence of both time-integrated and time-resolved PL, presented in figure\,\ref{fig:fig2}(d) and (e),
respectively.  The intensity of the I$_\mathrm{X}$  transition initially increases with increasing temperature reaching a maximum at $\simeq 30$\,K. Above this temperature, the PL intensity decreases and can be hardly observed above 100K. %vanishing above $\simeq 140$\,K. 
The increase of the PL intensity with temperature confirms the indirect nature (in the momentum space) of the transition. A deeper insight into the
anomalous temperature dependence of the emission peak of I$_{\text{X}}$ is provided by time-resolved spectroscopy. Our
measurements show that the decay times of both fast and slow components decrease with increasing temperature (see figure
\ref{fig:fig2} (b) and figure S4). However, the intensities related to the fast and slow components (calculated as an integral of
individual exponentials) behave differently with the temperature. The integrated intensity of the fast component (momentum-direct transition)
decreases monotonically with increasing temperature, while the intensity of the slow component (momentum-indirect transition) initially
increases with increasing temperature, displaying a similar trend to that observed for the time integrated PL intensity (see
figure \ref{fig:fig2}(d)). This suggests that the initial increase of the PL intensity as a function of the temperature can be
attributed to the increasing efficiency of the indirect transition with the increasing population of phonons. The decrease in
intensity at higher temperatures is a result of the thermal activation of non-radiative recombination channels, which eventually dominate over the recombination process. The exact nature of the non-radiative process is not known. The activation energy for the photoluminescence quenching is about 30meV (see supplementary information fig. S6), which suggests a contribution from defects states rather than the recently proposed interlayer-exciton dissociation\cite{Mouri2017} (the activation energy for this process was found to be $\sim$90meV).

The inter-layer nature of the I$_\mathrm{X}$  emission is further confirmed by the photoluminescence excitation (PLE)
spectroscopy presented in figure\,\ref{fig:fig3}(a). The inter-layer emission vanishes if the excitation energy is below that of
excitonic states in MoSe$_2$ (see also fig. S3(b)). A clear enhancement of the I$_\mathrm{X}$ peak is observed when the energy of the excitation is
resonant with A- and B-excitons of MoSe$_2$ and A-exciton of MoS$_2$. We do not observe any clear spectral feature at energies corresponding
to the B-exciton of MoS$_2$, which might be related to the opening of new relaxation paths for high energy excitons\cite{Kozawa2014}.

The PL energy of the inter-layer exciton compared to MoSe$_2$ and MoS$_2$ intra-layer excitons provides a lower bound for
the conduction and valance band offsets, which we estimate to be  $\sim$300\, meV and 550\, meV, respectively. Both these values
are smaller by $\sim$100\,meV than theoretically predicted offsets \cite{Kang2013,zhang2016systematic}, which indicates on the reduction of
inter-layer exciton binding energy as expected for spatially separated carriers.

\section{Circularly polarized emission of the inter-layer exciton}

\begin{figure*}[t!]
\centering
\includegraphics[width=1\linewidth]{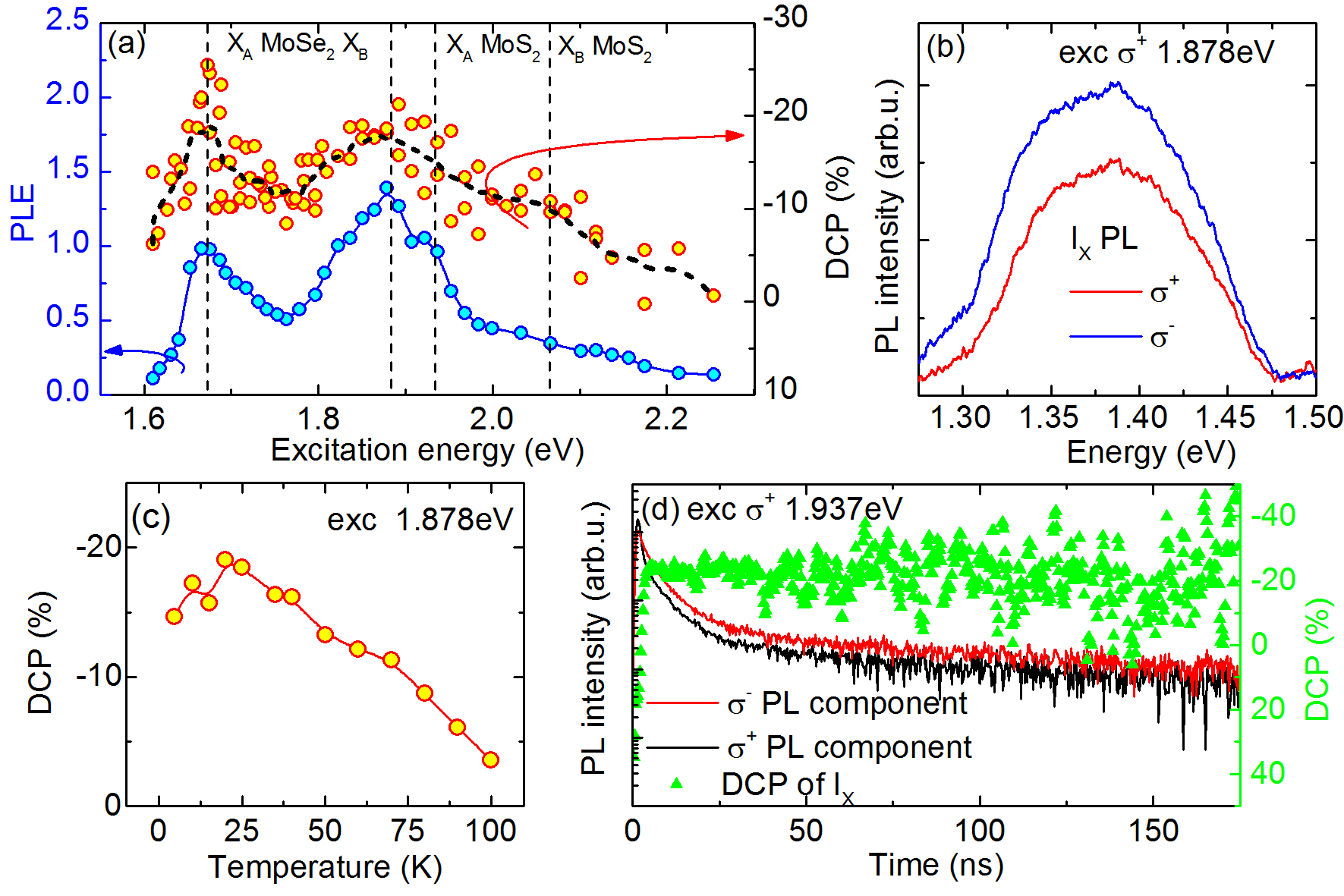}
\caption{(a) PLE spectra of the I$_\mathrm{X}$ transition at $T=5$\,K (blue points). Dashed vertical lines indicate the transition
energies of A- and B-excitons in single layer MoSe$_2$ and MoS$_2$. Orange points show dependence of the degree of circular
polarization versus the excitation photon energy (dashed thick black line is average value of DCP). (b) Circular polarization resolved PL spectrum of I$_\mathrm{X}$
 under $\sigma^+$ excitation. (c) Temperature dependence of the degree of circular
polarization of I$_\mathrm{X}$. (d) Comparison of $\sigma^+$ (black line) and $\sigma^-$ (red line) of I$_{\text{X}}$ PL decay
component and resulting degree of polarization .} \label{fig:fig3}

\end{figure*}

When using circularly polarized excitation the I$_\mathrm{X}$ emission is partially circularly polarized with the opposite
helicity, as shown in figure \ref{fig:fig3}(b). This is in strong contrast with  the co-polarized emission of the A-exciton in
MoSe$_2$ \cite{Surrente2017}. In addition, under linearly polarized excitation, which simultaneously excites both
valleys, I$_{\text{X}}$ does not exhibit any detectable degree of circular polarization, which rules out a ``built in'' imbalance
of the valley population \cite{Scrace2015} or carrier lifetimes as the origin of the observed polarization behavior. The opposite
helicity compared to the excitation points to a more complex valley and spin transfer mechanism than that previously invoked to
explain the non-trivial dependence of the polarization of inter-layer excitons in gated heterostructures \cite{Rivera2016}.

Since the pump-probe experiments indicate a very fast inter-layer exciton/charge transfer \cite{Ceballos2014, Chen2016}, we might
expect that the radiative recombination of inter-layer exciton give PL co-polarized with the excitation (as in case of monolayers
and WSe$_2$/MoSe$_2$ heterostructure\cite{Rivera2016}). The observed counter polarized PL indicates the influence of additional
scattering processes resulting in a reversed hole valley index during inter-layer exciton formation (since the electrons are scattered very effectively between valleys \cite{Hsu2015,molina2017ab}, they cannot determine observed polarization). It is worth noting that
a partially different alignment of bright and dark excitonic states \cite{Zhang2015,Baranowski2017,Echeverry2016} in each layer of
the MoSe$_2$/MoS$_2$ and MoSe$_2$/WSe2$_2$ heterostructures suggests that dark excitons may play a role in the observed
polarization of the inter-layer exciton. Nevertheless, a more profound understanding of the heterostructure band structure and
inter-layer exciton formation is needed to fully account for the counter polarized emission of inter-layer exciton.

It is interesting that the opposite degree of PL polarization (see figure \ref{fig:fig3}(a)) depends on the excitation wavelength
and is enhanced for resonant excitation of the A and B excitons in MoSe$_2$  and the A exciton in MoS$_2$. This shows that the
degree of circular polarization of the inter-layer exciton reflects the initial degree of valley polarization of the photo
created carriers. The polarization of inter-layer excitons is expected to be quite robust due to the reduced overlap of the
electron-hole wave functions, which minimizes the exchange interaction. Therefore, the observed degree of circular polarization
depends on the depolarization of hot intra-layer excitons before they relax to the inter-layer state. Since inter-valley
scattering is very efficient for intra-layer excitons with high kinetic energy (the inter-valley scattering for 100\,meV detuning
for MoS$_2$ is in the range of few tens of fs\cite{Yu2014}, comparable with the inter-layer charge transfer time scale
\cite{Ceballos2014, Chen2016}) the out of resonance excitation leads to a reduced polarization of hot inter-layer excitons and
therefore to a reduced polarization of the inter-layers excitons.

The degree of circular polarization of the I$_\mathrm{X}$ emission as a function of temperature is shown in figure
\ref{fig:fig3}(c). Initially, it increases with increasing temperature, which is probably associated with the initial rapid
decrease of the radiative lifetime seen in figure \ref{fig:fig2}(d). At higher temperatures ($\geq25$\,K) the degree of circular
polarization decreases, possibly due to the thermal activation of ``co-polarized'' emission channels.

In figure\,\ref{fig:fig3}(d) we show the temporal evolution of polarization resolved PL. A stronger signal for the cross
polarized component of PL is observed over more than 100\,ns. The measured degree of circular polarization ($\simeq 20$\%) does
not decrease within the available time window (175\,ns) in which the signal to noise ratio allows us to reliably determine the
degree of circular polarization. The extremely long lived polarization demonstrates that intervalley scattering is strongly
suppressed for the inter-layer exciton, which is very promising for future valleytronic applications.

\section{Conclusions}
We have observed inter-layer exciton emission from MoS$_2$/MoSe$_2$/MoS$_2$ trilayer stacks.We shown that our fabrication approach, based on CVD grown flakes sequentially transferred over a large area, is capable of providing high quality van der Waals heterostructures
with strong potential for scalability. Our systematic studies of temperature and temporal dependence of the inter-layer PL reveal the degeneracy of momentum-direct and indirect band gaps in the investigated heterostructure. Under circularly polarized excitation, we observe
long lived, counter polarized  PL of the inter-layer exciton. This effect provides a new and previously unsuspected degree of
freedom for tailoring the properties of van der Waals heterostructures. The degree of circular polarization of inter-layer
emission is stable over hundreds of ns, which is promising for future applications.

\section*{Methods}

MoS$_2$/MoSe$_2$/MoS$_2$ stacks have been prepared by two separate transfer steps using a wet transfer KOH method
\cite{Wang2016Hetero}. The layer of MoS$_2$ flakes was first transferred onto the as-grown MoSe$_2$ single layer film on
sapphire. Next, the MoS$_2$-MoSe$_2$ stack was transferred onto an as-grown lower-MoS$_2$ monolayer film on a sapphire substrate
\cite{Surrente2017}.

The PL measurements were performed using a CW frequency doubled solid state laser emitting at 532\.nm. PLE spectroscopy was
performed using a femtosecond pulse (150 fs) mode-locked Ti:sapphire laser or frequency doubled output of an optical parametric
oscillator (OPO), synchronously pumped by the Ti:sapphire laser. The excitation beam was focused on the sample by a $50\times$
microscope objective with a numerical aperture of 0.55, giving a spot size of approximately 1$\mathrm{\mu m}$. The emitted PL was
collected through the same objective and redirected to a spectrometer equipped with a liquid nitrogen cooled CCD camera. Time
resolved measurements were performed in the single photon-counting mode. The emission from the samples was spectrally filtered by
a bandpass filter and the signal was detected using a Si avalanche photodiode.

\begin{suppinfo}
Interlayer exciton photoluminescence mapping; micro-graph showing border between as-grown material and transfer zone; photoluminescence power dependence for two more places in the transfer zone; time resolved photoluminescence signal outside the transfer zone and excited below MoSe$_2$ ground state; inter-layer exciton photoluminescence decay as a function of temperature; Arrhenius plot of inter-layer exciton photoluminescence intensity.    
\end{suppinfo}

\begin{acknowledgement}
This work was partially supported by ANR JCJC project milliPICS, the R{\'e}gion Midi-Pyr{\'e}n{\'e}es under contract MESR
13053031, BLAPHENE and STRABOT projects, which received funding from the IDEX Toulouse, Emergence program,  ``Programme des
Investissements d'Avenir'' under the program ANR-11-IDEX-0002-02, reference ANR-10-LABX-0037-NEXT, and by the PAN--CNRS
collaboration within the PICS 2016-2018 agreement. N. Z.\ holds a fellowship from the Chinese Scholarship Council (CSC). This
work was financially supported by the Swiss SNF Sinergia Grant no. 147607 and Marie Curie ITN network MoWSeS (grant no. 317451).
K.W. and S.M. acknowledge support from DEC-2013/10/E/ST3/00034 funded by the National Science Center of Poland.
\end{acknowledgement}

\emph{The authors declare no competing financial interest.}

\bibliography{bib_indirect}

\end{document}